\documentstyle[12pt]{article}

\font\sqi=cmssq8
\def\DR{\rm I\kern-1.45pt\rm R}
\def\DC{\kern2pt {\hbox{\sqi I}}\kern-4.2pt\rm C}

\begin{document}
{\hfill physics/9712027}
\begin{center}
{\Large\bf Anyons, Monopole and Coulomb Problem}
 \\
\vspace{0.5cm}
{\large A.Nersessian}\footnote{e-mail:nerses@thsun1.jinr.dubna.su},
{\large V.M.Ter-Antonyan}\footnote{e-mail:terant@thsun1.jinr.dubna.su}
\vspace{0.5cm}
\\
{\it Bogolyubov Laboratory of Theoretical Physics, JINR} \\
{\it Dubna, Moscow region, 141980, Russia}
 \end{center}
\begin{abstract}
The monopole systems with  hidden symmetry of the
two-di\-men\-si\-onal Coulomb problem are considered.

One of them, the ``charge-charged magnetic vortex" (``charge-$Z_2$-dyon)
with a  half-spin, is constructed by reducing
 the quantum circular oscillator with respect to the action of
the parity  operator.

The other two systems are constructed by reduction from
 the two-di\-men\-si\-onal
complex  space. The first system is a particle on the sphere
in the  presence of the exterior constant magnetic field
(generated by  Dirac's monopole located in its center).
This system is dual to the massless (3+1)-di\-men\-si\-onal particle
with  fixed energy.
The second system represents the particle on the pseudosphere in the
presence of exterior magnetic field and is dual to the massive
 relativistic anyon.
\end{abstract}
\setcounter{page}1
\setcounter{footnote}0
 \setcounter{equation}0
\section{Introduction}
As it is well-known, most of the integrable systems of classical mechanics
can be constructed by the Hamiltonian reduction of
the simplest systems of the type of a free particle or an oscillator
formulated on a  larger phase space \cite{perelomov}.
This kind of construction not only classifies the known integrable systems
but also allows one to construct new integrable systems together with
their explicit solutions.

The procedure of reduction becomes especially important  in quantum
mechanics where it is much more difficult to establish integrability
 and exact solvability of  systems.
For example, thought any one-di\-men\-si\-onal system of classical mechanics is
integrable and  exactly solvable, the corresponding quantum-mechanical system
being integrable is exactly solvable only in exceptional cases.

On the other hand, the Hamiltonian reduction provides a
 natural framework to
construct integrable {\it spinning} systems,
including multiparticle integrable  ones (see, e.g., \cite{arut}).

An interesting feature is that   such
two-, three- and higher di\-men\-si\-onal  systems  can be interpreted as
the ones interacted with  external gauge fields of monopoles.
Correspondingly, the spin in these systems can be viewed as a
consequence of the  nontrivial topology
 induced by the presence of a monopole.
Moreover,   in the dual interpretation, i.e. under exchange of the
role of coordinates  and momenta,  these systems can be interpreted as
free relativistic systems of spinning particles in an appropriate
gravitational background.

A good illustration is a Hamiltonian reduction of the four-di\-men\-si\-onal
isotropic  oscillator under the $U(1)-$group action with the nonzero
value $s$ of the Hamiltonian generator  of that group.

As a result of that reduction, the oscillator phase space $T_*\DC^2$
 reduces  to the phase space of a nonrelativistic
charge-Dirac monopole system which is $T_*\DR^3$
  with the symplectic structure
\begin{equation}
dp_a\wedge dq^a +s\varepsilon_{abc}\frac{q^a}{|q|^3}dq^b\wedge dq^c,
\quad a,b,c=1,2,3
\label{0}\end{equation}
and the rotation generators
\begin{equation}
J^a=\varepsilon^{abc}p_bq_c+s\frac{q^a}{|q|}.
\label{01}\end{equation}
The energy levels of the oscillator  reduce to the negative energy
levels of   the nonrelativistic ``charge-dyon" system described
by the Hamiltonian \cite{zwanziger}
\begin{equation}
   H=\frac{p^2}{2\mu}-\frac{s^2}{2\mu|q|^2}+\frac{\alpha}{|q|}
\label{chdyon}\end{equation}
Notice that  the ``centrifugal" term in this Hamiltonian
 provides the interaction of the induced dipole momentum
of the systems with the monopole magnetic field \cite{tn}.

In the dual picture, i.e. after the exchange of the role of the coordinates
and momenta, the reduced phase space describes (3+1)-di\-men\-si\-onal
massless particles with the helicity $s$.
This duality becomes clear in a four- di\-men\-si\-onal picture where the
prototypes of  phase space variables and the generators of spatial rotations
define the well-known  twistor realization of the (3+1)-di\-men\-si\-onal
Poincar\'e group for massless particles.

  When the initial system  reduces with respect to Hamiltonian action of
the non-Abelian group , the resulting system acquires isospin
degrees of freedom.
An example is the generalization of the five-di\-men\-si\-onal Coulomb problem,
which includes the interaction with  the $SU(2)$ monopole field.
This system was constructed by  reducing the eight-di\-men\-si\-onal isotropic
oscillator   under the  $SU(2)$-group action both in the classical \cite{iwayi}
and quantum \cite{leva} cases.

In the general case, the system obtained by   reduction under
 Hamiltonian action of the  $G$ group,
posesses an  interaction with an external
$G$-field, and the values of the generators of the $G$-group play the role of
charges \cite{stern}.

In the present paper, we consider the above mechanisms
on the systems which are related to the two-di\-men\-si\-onal Coulomb problem.

In  {\it Section 2},  we show that despite the classical equivalence
of the two-di\-men\-si\-onal Coulomb problem and circular oscillator,
established by the Bohlin-Levi-Chivita transformation, the quantum
correspondence  between these systems requires  reduction of
the quantum circular oscillator by the parity operator
($Z_2$-group) action. As a result, the oscillator breaks into two systems,
the two-di\-men\-si\-onal Coulomb problem, and its generalization, with
half-spin induced by the magnetic vortex (or, equivalently,
two-di\-men\-si\-onal ``charge-$Z_2$-dyon" system) \cite{bohlin}.

In {\it Section 3},  we perform the Hamiltonian reduction of
four- di\-men\-si\-onal systems  to particles on the sphere and pseudosphere
in  homogeneous magnetic fields.

The first system is a particle on a sphere which interacts with a
magnetic monopole located in its center, and it is dual  to the
(3+1)-di\-men\-si\-onal massless particle with fixed energy.
The second system has a dual interpretation as a free massive
relativistic anyon \cite{anyon}.

 \setcounter{equation}0
\section{Coulomb Problem with  Magnetic Vortex}

Consider the correspondence between the two-di\-men\-si\-onal Coulomb
problem and classical two- di\-men\-si\-onal isotropic oscillator
(circular oscillator) \cite{Bohlin}.

It is convenient to describe the Hamiltonian system,
 corresponding to the circular oscillator,
in terms of the phase space $T_*\DC$ by using
 the following canonical Poisson brackets and Hamiltonian
\begin{equation}
\{\pi, z\}=1,\quad\{{\bar\pi},{\bar z}\}=1;\quad
H_{osc}=\omega(\pi\bar\pi+ z{\bar z}).\end{equation}

This system has three  integrals of motion
which forms the  $su(2)$ algebra:
the oscillators  rotational momentum   $J$
and the two other generators which  are convenient to be presented
 in the form of
the complex (vector)  integral of motion $I^+$
and its complex-conjugated one $I^-$:
\begin{equation}
 J=i(\pi z-\bar\pi\bar z),\quad I^{+}=\omega({\pi^2}-{\bar z}^2)\;:
\quad \{I^+,I^-\}=2\omega^2J, \;\{I^{\pm},J\}=\pm 2I^{\pm}
\end{equation}

Let us exclude the origin of the coordinates of $\DC$,
and perform the transformation $(z,\pi)\to (w,p)$, which is canonical
on the  space  $T_*{\dot{\DC}}$
(and singular in the origin of the coordinates of $\DC$):
\begin{equation}
w= z^2 ,p=\pi/2 z:\;\;\; \{w,w\}=0,\;\{w,p\}=1,\; \{p,{\bar p}\}=0.
\label{bo}\end{equation}

This transformation, after rescaling $w\to w{\sqrt{2\mu\omega}}$,
$p\to p/{\sqrt{2\mu\omega}}$, converts the energy levels of the oscillator
$H_{osc}=E_{osc}$ into those of the Coulomb problem  $H_{C}=-2\mu\omega^2$
where
\begin{equation}
H_C=\frac{1}{2\mu}p{\bar p}-\frac{E_{osc}}{|w|};\quad
\end{equation}
 The oscillator  integrals of motion are transform into the rotational
momentum and Runge-Lenz vector of the two-di\-men\-si\-onal
Coulomb problem
\begin{equation}
 2{\tilde J}=2i\left( wp-{\bar w}{\bar p}\right),\quad
  {\tilde I}=\frac{i(J p+pJ)}{2}-\frac{E_{osc} w}{4|w|}.
\label{ik}\end{equation}
Note that in passing from the oscillator to the Coulomb problem
the rotational
momentum  of the system is doubled.
This reflects the fact, that  the single round of the orbit
in the oscillator problem corresponds to the double round of
the orbit in the Coulomb one.
To verify this, we parametrize the oscillator trajectories by
Zhukovski's ellipse $z=u+1/u$ where the complex parameter $u$
parametrizes a circle with a radius different from unity,
 $|u|=const\neq 1$.
Then, as a result of transformation $w=z^2$ we have
$$z=u+1/u \rightarrow w=u^2+1/u^2 +2,$$
which means that  the center of attraction of the Coulomb problem
is shifted to the focus of ellipse,
and a single round of the ellipses by the particle
in the oscillator problem  corresponds to the double rounding
of the orbit in the Coulomb one.

So, when we deal with elliptic orbits, we can ascertain the equivalence
 of the   circular oscillator and two-di\-men\-si\-onal Coulomb
problems on the classical level. \\

In the corresponding quantum problem we have
\begin{equation}
H_{osc}({\hat\pi},{\hat{\bar\pi}}, z,{\bar z})\Psi
=E_{osc}\Psi,
\quad \Psi(|z|,{arg\; z})= \Psi(|z|,{arg\; z} +2\pi),\end{equation}
 where ${\hat\pi}=-i\hbar{\partial}_{z}$, ${\hat{\bar\pi}}=
-i\hbar{\partial}_{\bar z}$.

Though transformation (\ref{bo}) turns the
Schr\"odinger equation of the oscillator into one
for the two-di\-men\-si\-onal Coulomb problem,  it breaks the single-valuedeness
requirement   for the wave functions,
$ \Psi(|w|, {\rm arg}\; w + 4\pi)=
\Psi(|w|, {\rm arg}\; w )$,
i.e. the quantum circular oscillator problem transform into
 the Coulomb problem on the {\it two-sheet Riemann surface}.

Therefore, we should supply the transformation (\ref{bo})
with the reduction of the oscillator  Schr\"odinger equation
by the  $Z_2$-group action given by  the parity operator.
This means that we may restricts ourselves to the even
 ($\sigma=0$) or odd  ($\sigma=\frac 12$) solutions of the oscillator
Schr\"odinger equation
 \begin{equation}
  \Psi_\sigma (z, \bar z) =
\psi_{\sigma} (z^2, {\bar z}^2){\rm e}^{i\sigma {\rm arg}\;z}.
\label{3}\end{equation}
Then, the wave functions  $\psi_\sigma$
obey the condition
$$
\psi_{\sigma}(|w|, {\rm arg}\; w + 2\pi)=
\psi_{\sigma}(|w|, {\rm arg}\; w ) ;
$$
 therefore, the domain of definition  ${\rm arg}\; w\in[0,4\pi)$
can be reduced, without loss of generality, to ${\rm arg}\; w\in[0,2\pi)$.

This procedure results in the Schr\"odinger equation
\begin{equation}
(\frac{1}{2\mu}{\hat p}_\sigma {\hat p}^+_\sigma-
\frac{\alpha}{|w|} +{\cal E} )\psi_\sigma=0,
 \label{4}\end{equation}
 where ${\cal E}=\mu\omega^2/{8}, \quad \alpha=E/{4}$,
and the momentum operator is defined by the expression
\begin{equation}
{\hat p}_\sigma = -2i\hbar\partial_w-\frac{i\hbar\sigma}{2w}.
\label{psigma}\end{equation}

 Substitution of  (\ref{psigma}) into (\ref{ik}) gives us the
expressions for the integrals of motion of the reduced system.

So, for both the values of $\sigma$, the obtained system
is characterized by the symmetry of the two di\-men\-si\-onal Coulomb problem.

From the eigenvalues of the oscillator rotational momentum $M$ and energy
 $E$
\begin{equation}
E=\hbar\omega(2N_r+ |M|+1),\quad
M= 0,\pm 1,\pm 2,...\quad  N_r=0, 1, 2,...,
 \end{equation}
we immediately derive the eigenvalues for the rotational momentum $m_\sigma$
and for the energy ${\cal E}^s$of the reduced system
\begin{equation}
{\cal E}^s=-\frac{\mu\alpha^2}{\hbar^2(N_r+ |m_\sigma|+1/2)^2} ,
\quad m_\sigma= \pm\sigma, \pm (1+\sigma), \pm (2 +\sigma),...
  \label{spec}\end{equation}
The wave
functions  of the
reduced system can easily be obtained from  the oscillator ones as well.

The Schr\"odinger equation of the reduced system
describes the motion of  spinless particle with the electric charge
 $e$  in the electric and magnetic fields with the potentials
\begin{equation}
 \phi= -\frac{\alpha}{e|w|},\quad  A_w=\sigma\frac{i\hbar c}{2ew},
\label{Aphi}\end{equation}
i.e. in the field of $Z_2$-dyon.
Indeed,  the magnetic potential   $A_w$ defines the magnetic field with
zero strength  $B=rot A_w=0$  ($w\in\dot{\DC}$) and constant magnetic flux
 $\sigma\pi\hbar c/2e$.

In other words, in the case of  $\sigma =0$
we have the two-di\-men\-si\-onal hydrogen atom, whereas for $\sigma=\frac 12$
we have the``charge-charged magnetic vortex"
(``charge-$Z_2$-monopole") system.

As we see from (\ref{spec}), the second system possesses spin $1/2$ and
the Aharonov-Bohm effect is observed in it.
      Note that the Hamiltonian  of the system
  (\ref{4})  can be represented in the form analogous to the one for
three-di\-men\-si\-onal ``charge-Dirac dyon"  system (\ref{chdyon})
$${\hat H}_\sigma=-\frac{4\hbar^2}{2\mu}{\partial_w}{\bar\partial}_w
+\frac{\hbar^2\sigma^2}{2\mu|w|^2} -\frac{\alpha}{|w|}.$$
\\

In a similar way , we can consider the transformation $w=z^N$
corresponding to the reduction by $Z_N$-group.
As a result of that reduction, the initial two- di\-men\-si\-onal system with
the potential  $|z|^a$ splits into
$N$ ``charge-magnetic vortex" (or ``charge- $Z_N$-monopole")
bound systems
 with the spin $\sigma=0, 1/N, 2/N,...(N-1)/N$
 and potential   $|w|^b$, where
 $ (a+ 2)(b + 2)=4 $, $N=1+a/2$.

 \setcounter{equation}0
\section{Particle on Sphere and Pseudosphere}

The hidden symmetry of the Coulomb problem establishes its trajectorial
correspondence with an isotropic oscillator  only in the two-, three-,
and five- di\-men\-si\-onal cases.

In the general case, the $n$-di\-men\-si\-onal Coulomb problem is
 trajectory-equivalent  to a geodesic flux on the
$n-$di\-men\-si\-onal sphere (attractive problem) or  $n$-di\-men\-si\-onal
pseudosphere
 (repulsive problem).
Integrals of motion of the Coulomb problem correspond to
the rotation   generators  on the (pseudo)sphere.
Equivalence  between these systems in the two-di\-men\-si\-onal case
can be established as follows:
 $$(1 +{pp^+})^2 w{w}^+ - mE=0 \;\;\Leftrightarrow \;\;
{pp^+}- \frac{{\sqrt{mE}}}{{\sqrt{ww^+}}}+1=0$$

It can be easily verified that the Hamiltonian of a particle on
a sphere (pseudosphere) under transformation (\ref{bo})
goes over into the squared Hamiltonian of a circular oscillator with
the positive (negative) coupling constant
$$ (1\pm {p{\bar p}})^2 w{\bar w}=(\pi{\bar\pi}\pm z{\bar z})^2.$$

Hereafter, we take advantage of the stereographic projection
of a sphere, $\DC P^1$ and Poincar\'e model for
pseudosphere (Lobachevsky plane) ${\cal L}$, where the metric is defined
by the expression
\begin{equation}
g(p, p^+)=\frac{mdpd{p}^+}{(1+ {pp^+})^2},
\label{fs}\end{equation}
 where $p^+={\bar p}$ for $\DC P^1$  and $p^+=-{\bar p}$ for ${\cal L}$.

Notice that  $\DC P^1$ and ${\cal L}$ can be constructed
by the Hamiltonian reduction of  $\DC^2$ and $\DC^{1.1}$
equipped by the K\'ahler structures $d\pi^\alpha d{\bar \pi}_\alpha$
under the $U(1)$-group action which is defined by the generator
$P=\pi^\alpha{\bar\pi}_{\alpha}$  with the value $P=m$.
Hereafter, indices are raised and lowered by using  the metric
$\eta_{\alpha{\bar\beta}}$ where
$\eta_{\alpha{\bar\beta}}=\delta_{\alpha{\bar\beta}}$ on $\DC^2$  and
 $\eta_{\alpha{\bar\beta}}=\sigma^3_{\alpha{\bar\beta}}$
on $\DC^{1.1}$.

The canonical Poisson brackets on  $\DC^2$ and $\DC^{1.1}$
are reduced to those  corresponding to the K\"ahler structures
 (\ref{fs}) under the following choice of  coordinates of
the reduced phase spaces
 $p_{(0)}=\pi^1/\pi^0$, ${\bar p}_{(0)}={\bar \pi}_1/{\bar \pi}_0$
 in the domain $\pi^0\neq 0$, and
 $p_{(1)}=\pi^0/\pi^1$, ${\bar p}_{(1)}={\bar \pi}_0/{\bar \pi}_1$, in the
domain $\pi^1\neq 0$.
So, a sphere can be covered with two charts with the transition functions
 $p_{(0)}= 1/p_{(1)}$, where $p$ parametrizes the complex plane,
while a pseudosphere can be covered with only one chart
parametrized by the coordinate $p_{(0)}$, $|p|< 1$  at $m< 0$
($p_{(1)}$, $|p|> 1$, at $m>0$).

Now consider the phase spaces  $T_*\DC^{2}$ and
$T_*\DC^{1.1}$, equipped with the canonical Poisson brackets
 \begin{equation}
\{\pi^\alpha, \omega_{\beta}\}=\delta^\alpha_\beta,\quad
\{{\bar\pi}^{\bar\alpha}, {\bar\omega}_{\bar\beta}\}=
\delta^{\bar\alpha}_{\bar\beta},   \quad \alpha,\beta =0,1 .
 \end{equation}

Let us carry out the Hamiltonian reduction of these spaces
by the action of the  generators
\begin{equation}
P=\pi^\alpha{\bar\pi}_{\alpha},\;
\;J=\frac{i}{2}(\pi^\alpha\omega_{\beta}-
{\bar\omega}^{\bar\alpha}{\bar\pi}_{\bar\alpha}),\quad \{P,J\}=0,
\label{full}\end{equation}
with the values
\begin{equation}
J=s, P= m.\;\;
 \label{ae}\end{equation}
Due to commutativity of these generators, reduced phase spaces are
four-di\-men\-si\-onal.
The following functions $(p, w)$ can be chosen as  local complex
coordinates  of the reduced phase space  (when $\pi^0\neq 0$)
\begin{equation}
 p=\pi^1/\pi^0,\quad
w=ig(p,\bar p)({\lambda}^1- p{\lambda}^{0}), \quad
\lambda^\alpha=\omega^{\alpha}/{\bar\pi}^0,
\end{equation}
where the metric $g(p,\bar p)$ is defined by  expression  (\ref{fs}).

The reduced Poisson brackets on these spaces are of the form
\begin{equation}
\{p,w \}=\{{\bar p},{\bar w}\}=1,\quad \{{w},{\bar w}\}=
i\frac{s}{m} g(p,{\bar p}).\label{rb}
\end{equation}

Thus, we have reduced the canonical Poisson brackets
on $T_*\DC^2$ and $T_*\DC^{1.1}$  to the
twisted ones on $T_*\DC P^1$ and ${\cal L}$, respectively.

We have got the phase spaces for the particle  on the sphere and
 pseudosphere in the  presence  of the constant homogeneous magnetic field.

For the sphere this  field is generated by the magnetic monopole
 located  in the center of the sphere.

Now let us define on  $T_*\DC^2$ and $T_*\DC^{1.1}$  the generators
\begin{equation}
P^a=\pi^\alpha T^a_{\alpha\bar\beta}{\bar\pi^{\bar\beta}}
\quad J^{a}=\frac{1}{2}T^a_{\alpha\bar\beta}
(\pi^\alpha {\bar\omega^{\bar\beta}}+ {\omega}^\alpha {\bar\pi^{\bar\beta}})
\label{5},\end{equation}
which form the algebra
\begin{equation}
\{P^a, P^b\}=0,\quad \{P^a, J^b\}=\varepsilon^{abc}P_c,
\{J^a, J^b\}=\varepsilon^{abc}J_c ,
\label{2+1}\end{equation}
 where  $T^a=\sigma^a$ for  $\DC^2$ and
$T^a=(\sigma^0,\sigma^1,\sigma^2)$ for  $\DC^{1.1}$;
 the vector indices  $a,b,c$ are raised and lowered by using
 the metric $g_{ab}=\frac 12 tr( T^a\eta T^b\eta )$.\\
 Correspondingly, for $\DC^2$ we have $g_{ab}=\delta_{ab}$, and
 (\ref{2+1}) is the  $e(3)$ algebra
while for  $\DC^{1.1}$
we have $g_{ab}=diag{(1,-1,-1)}$, and  (\ref{2+1}) is
the (2+1)- di\-men\-si\-onal Poincar\'e algebra $iso(1.2)$.

The generators (\ref{5}) commute with $P$ and $J$,
and therefore, can be reduced on  $T_*\DC P^1$ and ${\cal L}$
where they  take the form
 \begin{equation}
J^a=V^a(p)w+{\bar V}^a({\bar p}){\bar w}+\frac{s}{m}P^a,
\quad
P^a=m\frac{T^a_{0{\bar 0}} + T^a_{1{\bar 0}}{\bar p} +
 T^a_{0{\bar 1}}p + T^a_{1{\bar 1}}p{\bar p}}{1+p{\bar p}}.
 \label{pr}\end{equation}
with  $V^a(p)=i\partial_p P^a(p,{\bar p})$.

So, the system with  such a phase space possesses the own angular momentum
(``spin") $s^a=\frac{s}{m}P^a$.

The Hamiltonian of the free particle on the sphere (pseudosphere)
is related with $J^a$ by the expression
\begin{equation}
  H=g^{-1} w{\bar w}=J^a J_a -s^2,
\end{equation}
i.e. can be considered as a reduced spherical top.

Performing the Hamiltonian reduction of the initial phase space only by
the action  of $J$, we get the six-di\-men\-si\-onal phase space.
Thus, choosing the coordinates of the reduced phase space of the form
$$ P^a=\pi T^a{\bar\pi},\quad Q^a=i(\omega T^a\pi-\pi T^a\omega)/(2{\pi\bar\pi}),$$
we  see
 that $T_*\DC^2$ ($T_*\DC^{1.1}$)
is reduced to   $T_*\DR^3$ ($T_*\DR^{1.2}$) while the reduced symplectic
structure  and the reduced generators $J^a$ take the form
(\ref{0}) and (\ref{01})
after canonical  transformation $(Q^a, P^a)\to (P^a,-Q^a)$.

It is easy to see  that
$P^\mu=(P,P^a), M^{[\mu\nu]}=(J^a, (2{\pi\bar\pi})Q^a)$
define on  $T_*\DC^2$  the well-known twistor realization
of the (3+1)-di\-men\-si\-onal Poincar\'e group for the massless particles,
and  $J$ plays the role of their helicity.
This gives transparent explanation of the duality between
massless particles  and nonrelativistic ``charge-monopole" system
mentioned in the Introduction.

One can interpret the systems, reduced to sphere,
in term of the (3+1)-di\-men\-si\-onal Poincare group.
In this case, the radius of the sphere has a meaning of
the energy of the massless particle whereas the spin of the system
has a meaning of the helicity.

However, since the generator $P$  does not commute with the generators
of the  Lorentz rotation, the performed Hamiltonian reduction
to the sphere destroys the Lorentz invariance of the reduced system.\\

Unlike the system on the sphere, the one on the
pseudosphere admits complete relativistic interpretation \cite{anyon}.

Namely, the generators  $P^a, J^a$ define on $T_*\DC^{1.1}$
the action of the (2+1)-di\-men\-si\-onal Poincar\'e group and commute with
 $J$ and $P$. Thus, the reduced system is relativistic-invariant .

The ``Casimirs"  $ P^aP_a$ and $P_aJ^a$  satisfy the
following  conditions:
$$ P^aP_a=P^2,\;\; P_aJ^a=PJ.$$
 Hence, the generators  $J$ and $P$ can be interpreted as those
of spin and mass of the relativistic particle.
Note that the transformation $\pi^1\leftrightarrow\pi^0,
\omega^0\to\omega^1$, leads to  $(m, s)\to (-m,-s)$ and, consequently,
is an identity transformation in terms of the Poincar\'e group.
So, the Hamiltonian reduction by the action of the generators $P$ and $J$
fixes the mass $|m|$ and  spin $s$  of a free relativistic
two-di\-men\-si\-onal particle.\\

Let us now outline the scheme of an analogous reduction in the quantum case.
Introduce the operators

$$ {\hat{\bar\omega}}_{\alpha} ={\partial}/{\partial\pi^\alpha},\quad
{\hat\omega}^{\alpha} =-{\partial}/{\partial{\bar\pi}_\alpha},$$
and impose on the wave function $\Psi(\pi,{\bar\pi})$ the
conditions
  \begin{equation}
{\hat P}\Psi(\pi,{\bar\pi})= m\Psi(\pi,{\bar\pi}),
\quad {\hat J}\Psi(\pi,{\bar\pi}) = s\Psi(\pi,{\bar\pi})
\label{qr}\end{equation}
which are the quantum analogs of the constraints (\ref{ae}).

The solution to this system can be represented in the form
\begin{equation}
\Psi (\pi,{\bar\pi})=\psi_{m,s} (p, {\bar p}) e^{is\gamma(\pi,{\bar\pi})},
 \quad{\rm where}\quad {\hat J}\gamma= i,
\label{wfs}\end{equation}
from which it follows:
\begin{equation}
{\hat w}\Psi=g(p,\bar p)({{\hat\lambda}}^1- p{{\hat\lambda}}^{0})\Psi
 = e^{is\gamma} {\hat w}_{red}\psi(p,{\bar p}),
\end{equation}
where
\begin{equation}
{\hat w}_{red}=i\frac{\partial}{\partial p}+\frac{s}{2m}  A(p,{\bar p}),
\quad  A_{\gamma}(p,{\bar p})={\hat w}\gamma .
\end{equation}
So,  $\psi(p,\bar p)$ and ${\hat w}_{red}$  define
the wave function    and the momentum operator of the reduced system.

Choosing
$\gamma_+={i}\log{\pi^0}/{\bar\pi_0}$ or
 $\gamma_-={i}\log{\pi^1}/{\bar\pi_1}$, we  find
 $A_+(p,{\bar p})=i{\partial_p K(p,\bar p)}$
and $A_-(p,{\bar p})=i{\partial_{\bar p} K(p,\bar p)}$,
respectively, where  $K(p,{\bar p})=m\log(1+p{p}^+)$.

We can also choose   $\gamma=(\gamma_+ +\gamma_-)/2$.
In this case, each chart contains a singularity in  $p=0$,
while the
 choice  $\gamma_\pm$ at $p=p_{\pm}$ leads to a system,
being regular on each chart.

The reduction of the initial system by only generator
$J$ suggest the ansatz
 $\Psi(\pi,{\bar\pi})=\psi(P^a)e^{is\gamma}$
which results in the mometum operator with the
vector potential of the Dirac monopole at $\gamma=\gamma_{\pm}$,
and vector potential of the
Schwinger one at $\gamma=(\gamma_+ +\gamma_-)/2$.

It is clear that $\psi_{m.s}(p,\bar p)=\psi_s(P^a)$ where $P^a$
are defined by (\ref{pr}). \\

The choice of   $\gamma$   depends on a given problem.
For example, the requirement of invariance of the wave function on
the sphere under transition from one chart to another
implies the choice $\gamma=(\gamma_++\gamma_-)/2$ which corresponds
to the Schwinger monopole.
If  $\gamma=\gamma_{\pm}$, the reduced system is regular on both the charts,
i.e. we have Dirac's monopole.

The requirement for the wave function to be
 single-valued leads to the quantization of spin:
 it takes integer or half-integer values.

Now, consider the system on  $\DC^{1.1}$ corresponding
to that on $\DR^{1.2}$ and on the pseudosphere.

In this case the choice  $\gamma=\gamma_\pm$
leads to  systems with regular vector potentials on
both $\DR^{1.2}$ and  the pseudosphere
(since the later can be covered with one chart).
Therefore, the requirement of regularity does not impose any quantization
condition on $s$.
As we have shown, a system on a pseudosphere can be interpreted as
 a two-di\-men\-si\-onal free relativistic particle, that is an anyon,
because its spin is not quantized\cite{anyon},\cite{poly}.

One can look at this picture from  another point of view.
We have shown that
$(\pi^1,\omega^1)\leftrightarrow (\pi^0,\omega^0)$
is the identity transformation for a two-di\-men\-si\-onal
relativistic particle.
Taking  $\gamma=(\gamma_++\gamma_-)/2$, we obtain that the requirement
of the single-valuedness of the wave function $\Psi_s$
does not leads to the quantization of $s$ due to the signature of
 the metric on $\DC^{1.1}$.
On the other hand, the initial system with a given choice of $\gamma$
can be viewed as the system of two identical particles.
Thus,  interchanging these particles, we  find
that the wave function acquires the phase  $2\pi s$.

\section{Acknowledgments}
 The authors would like to thank G. Pogosyan for a given possibility
to report this work at the
VIII International Conference ``Symmetry Methods in Physics",
Dubna, 1997.

The work of one of the authors (A.N)
has been supported in part by grants
 INTAS-RFBR No.95-0829, INTAS-96-538, and INTAS-93-127-ext .

\end{document}